# Fabrication of a multifunctional photonic integrated chip on lithium niobate on insulator using femtosecond laser assisted chemo-mechanical polish

**Rongbo Wu** [1,2], **Jintian Lin** [1], **Min Wang** [3,4], **Zhiwei Fang** [3,4], **Wei Chu** [1], **Jianhao Zhang** [1,2], **Junxia Zhou** [3,4], **and Ya Cheng** [1,3,4,5],*

[1] State Key Laboratory of High Field Laser Physics, Shanghai Institute of Optics and Fine Mechanics, Chinese Academy of Sciences, Shanghai 201800, China; rbwu@siom.ac.cn (R.W.); jintianlin@siom.ac.cn (J.L.); chuwei@siom.ac.cn (W.C.); jhzhang@siom.ac.cn (J. Z.)

[2] University of Chinese Academy of Sciences, Beijing 100049, China

[3] State Key Laboratory of Precision Spectroscopy, East China Normal University, Shanghai 200062, China; mwang@phy.ecnu.edu.cn (M.W.); zwfang@phy.ecnu.edu.cn (Z.F.); 52180920026@stu.ecnu.edu.cn (J.Z.)

[4] XXL—The Extreme Optoelectromechanics Laboratory, School of Physics and Materials Science, East China Normal University, Shanghai 200241, China

[5] Collaborative Innovation Center of Extreme Optics, Shanxi University, Taiyuan 030006, Shanxi, China

* Correspondence: ya.cheng@siom.ac.cn

**Abstract:** We report fabrication of a multifunctional photonic integrated chip on lithium niobate on insulate (LNOI), which is achieved by femtosecond laser assisted chemo-mechanical polish. We demonstrate a high extinction ratio beam splitter, a 1 × 6 optical switch, and a balanced 3 × 3 interferometer on the fabricated chip by reconfiguring the microelectrode array integrated with the multifunctional photonic circuit.

## 1. Introduction

Photonic integrated circuits (PICs) pave an endless path for the future development of information technology. The high expectation is established on the facts that photons can efficiently transfer and process large amount of information at high speed. Once the photonic information processing can be conducted on integrated microchips in a scalable fashion, many aspects of our modern society will be revolutionized. Until recently, silicon photonics has been the main platform for PIC applications owing to its compatibility with the mature complementary-metal-oxide-semiconductor (CMOS) technology [1,2]. In the meantime, lithium niobate (LN) photonics has been emerging as an alternative and/or complementary approach for realizing large-scale PICs, thanks to the recent advances in the development of fabrication technologies of high quality photonic micro- and nanostructures such as low-loss waveguides and high quality factor (Q factor) microresonators on lithium niobate on insulator (LNOI) [3–12].

In this work, we take a crucial step toward realization of large-scale LNOI-based PICs by demonstrating a multifunctional photonic integrated chip. The idea is to take the advantage of reconfigurability as provided by the large electro-optic coefficient of crystalline LN which allows us to swiftly switch the optical paths in the LN PICs. The functionalities of the fabricated device are examined in a quantitative manner. It is noteworthy that our device is fabricated using femtosecond laser assisted chemo-mechanical polish, which is compatible with other optical lithographic technologies as we have explained before [8].

## 2. Materials and Methods

The photonic integrated chip was fabricated on an x-cut LNOI wafer with a thickness of 700 nm (NANOLN, Jinan Jingzheng Electronics Co., Ltd., Jinan, Shandong, China) using the technique described in great detail in [13]. Briefly speaking, the fabrication procedures include: (1) space-selective ablation of a chromium (Cr) layer coated on top of the LNOI to generate the pattern of a pre-designed PIC chip using a focused femtosecond beam; (2) chemo-mechanically polish of the sample (i.e., the side with the Cr coating); in this step, the LN without being protected by the Cr mask will be completely removed and the LN underneath the Cr mask will survive from the chemo-mechanical polish thanks to the high hardness of Cr; (3) removal of Cr coating by chemical wet etching; (4) a secondary chemo-mechanical polish to further eliminate the roughness near the top surface of the LNOI waveguides which results from the rough edge of the Cr mask patterned by the femtosecond laser ablation. The surface roughness achieved by the chemo-mechanical polish reached ~0.45 nm for all sides undergone the chemo-mechanical polish; (5) coating of the fabricated sample with a $Ta_2O_5$ layer of a thickness of 3.5 μm. The small refractive index contrast between LN (refractive index ~2.138 ($n_e$), 2.211 ($n_o$)) and $Ta_2O_5$ (refractive index ~2.058) ensures the waveguides to operate in the single-mode waveguiding regime; (6) depositing a thin layer of gold of a thickness of ~100 nm on the $Ta_2O_5$ layer; and (7) patterning the gold layer into microelectrodes by space-selective ablation with focused femtosecond laser pulses. In our experiment, the femtosecond laser ablation was conducted at a repetition rate of 250 kHz and a scan speed of 40 mm/s. The center wavelength of our femtosecond laser is 1030 nm, and the pulse width is 170 fs. The laser powers chosen for patterning the Cr and Au layers are slightly above the ablation threshold to ensure high-resolution microfabrication. An objective lens (Model: M Plan Apo NIR, Mitutoyo Corporation, Japan) with a numerical aperture (N.A.) of 0.7 was used to focus the laser pulses, creating a focal spot of a diameter of ~1 μm on the sample.

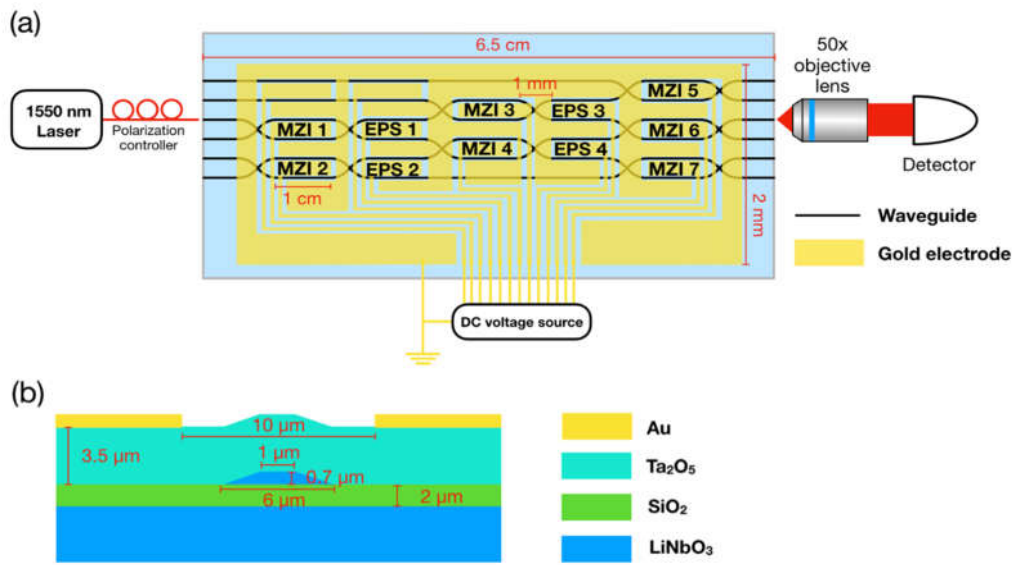

**Figure 1.** (a) Schematic of the multifunctional photonic integrated chip and the characterization system. (b) Cross sectional geometry of a LNOI waveguide covered with $Ta_2O_5$ and sandwiched between a pair of gold electrodes built on top of the $Ta_2O_5$ layer.

Figure 1a shows the design of our photonic integrated chip as well as the experimental layout for the device characterization. The chip consists of seven Mach–Zehnder interferometers (MZI1-MZI7), four external electro-optic phase shifters (EPS1-EPS4) and an array of microelectrodes, which are accommodated in an area of a footprint size of 6.5 cm × 0.2 cm. The beam splitting ratio of each MZI can be precisely controlled by adjusting the voltages applied on the microelectrodes fabricated along the two arms. The arm lengths of both the MZI and EPS are 1cm, resulting in a total device length of 6.5 cm. An array of twelve gold electrodes are fabricated near the arms of MZI and EPS for reconfiguring the device in real time. Figure 1b shows the cross sectional geometry of a LNOI waveguide covered with $Ta_2O_5$ and sandwiched between a pair of gold electrodes built on the $Ta_2O_5$ layer. In our experiment, typically a voltage of 9.7 V is required to realize a phase shift of π in the 1 cm long MZI arms.

A laser beam generated by a tunable laser (LTB-6728, Newport Corporation, Santa Clara, CA, USA; wavelength ~ 1550 nm) can be coupled into any of the six input ports of the fabricated device using a fiber lens mounted on a 6-axis nano positioning stage (Throlabs Inc., Newton, NJ, USA). The gold microelectrodes are connected to direct current (DC) voltage sources through a DC multi-contact wedge (MCW-26, GGB Industries Inc., Naples, Florida, USA). The polarization of the input light is controlled by an in line polarization controller (FPC561, Throlabs Inc., Newton, NJ, USA). The output signal is collimated using a 50× objective lens (M Plan Apo NIR, Mitutoyo Corporation, Kawasaki, Kanagawa, Japan) mounted on a 3-axis nano positioning stage, and directed into a high-sensitivity InGaAs power meter (S155C, Throlabs Inc., Newton, NJ, USA) for power measurement, as illustrated in Figure 1a.

## 3. Results and discussion

Figure 2a shows the picture of fabricated chip taken by a digital camera, which is placed near a 1 Renminbi (RMB) coin for comparison of the sizes. The total time of femtosecond laser patterning of the photonic integrated chip is approximately 2 hrs, which, however, can be shortened by improving our motion stage in the future. Figure 2b shows the close-up view image of the chip taken under an optical microscope, in which the waveguides, beam splitters, and the gold electrodes can all be clearly seen. Figure 2c shows the zoom-in graph of the area indicated by the red dashed box in Figure 2b taken using a scanning electron microscope (SEM).

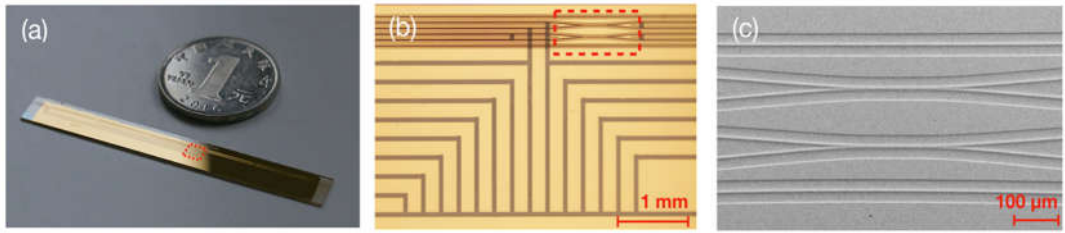

**Figure 2.** (a) Digital camera picture of the chip placed near by a 1 RMB coin. (b) Zoom-in micrograph of the area indicated by the red dashed box in (a). (c) SEM image of the beam splitters and waveguides before the coating of the gold layer, showing the smooth surface.

We demonstrate three functions of the device by taking the advantage of the reconfigurability, including a high extinction ratio cascaded MZI, a 1 × 6 optical switch, and a balanced 3 × 3 interferometer. These functionalities are fundamental for integrated photonic applications and have been demonstrated before in other material platforms such as silicon or fused-silica [14,15]. Therefore, we realize the same benchmark functionalities to evaluate the performance of the reconfigurable LNOI PIC fabricated with the chemo-mechanical polish. Below, we characterize the three devices one after another.

*3.1. High extinction ratio cascaded MZI*

First, we demonstrate a high extinction ratio cascaded MZI using only part of the photonic circuit as indicated by the waveguides highlighted in red in Figure 3a. To achieve the expected accurate beam splitting, we use an algorithm as described below:

1. Send the laser beam into port 1, and then tune MZI4 until the power of output 0 is minimized;
2. Tune EPS2 until the power of output 1 is minimized.
3. Simultaneously vary the electric voltages applied to MZI2 and MZI7 by the same quantity until the power of output 1 is minimized.
4. Tune EPS2 until the power of output 1 is maximized.
5. Simultaneously vary the electric voltages applied to MZI2 and MZI7 by the same quantity but of opposite signs until the power of output 1 is maximized.

6. Repeat steps 2-5 until the power of output 1 cannot be further reduced.

Figure 3b-c present the DC response of our cascaded MZI beam splitter, showing a half wave voltage of 9.7 V and an extinction ratio of ~28 dB. The demonstrated extinction ratio is comparable to the previous results reported in [14,15], and should be able to be improved by refining the experimental conditions. For instance, the current polarization controller used in our experiment may not be able to produce a TE wave of high purity, which may be the main cause of the spoiled extinction ratio of the beam splitter.

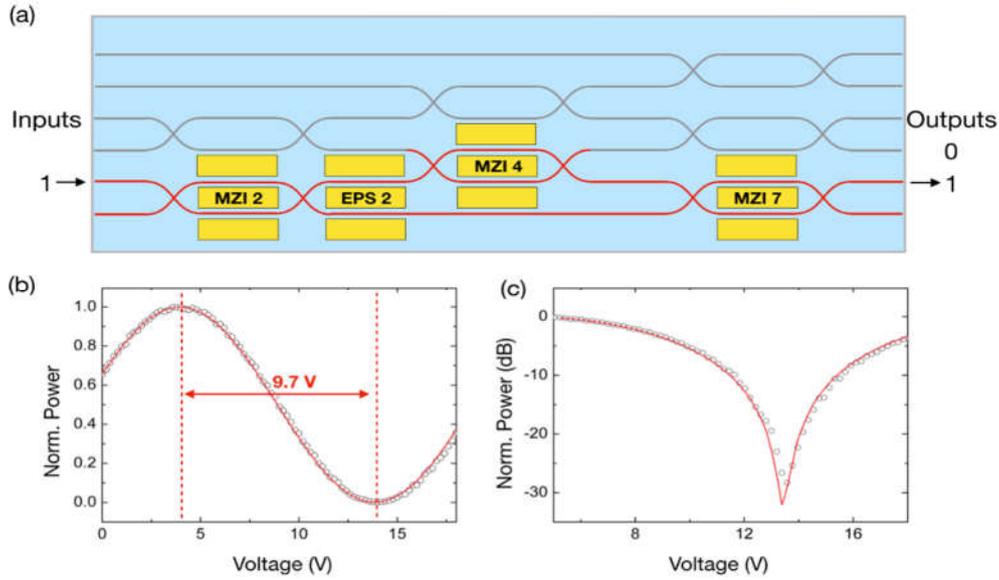

**Figure 3.** (a) Schematic of a high extinction ratio cascaded MZI beam splitter using part of the chip. The DC responses of cascaded MZI beam splitter in (b) linear and (c) logarithm scales when the laser is sent into input port 1 and collected from output port 1.

*3.2 1 × 6 optical switch*

Second, we demonstrate a 1 × 6 switch using the waveguides highlighted in red in figure 4 (a). We send the laser beam into either port 1 or 2, and then examine the output power of a predesignated output port such as output port 1. To direct the beam as much as possible to port 1, we sequentially optimize MZI1, MZI3, and MZI5, which are distributed along the path from input port 1 to output port 1, until the output power from port 1 is maximized. Likewise, the beam from input port 1 can be switched to the other output ports by sequentially optimizing the MZIs along the paths connecting the input port and the corresponding output port. Figure 4b and c show the normalized powers of the six output ports when the laser beam is sent into port 1 and 2, respectively. One can see that in all the measurements, almost 90% of the input power is recorded from the predesignated output port, indicating a high switching efficiency.

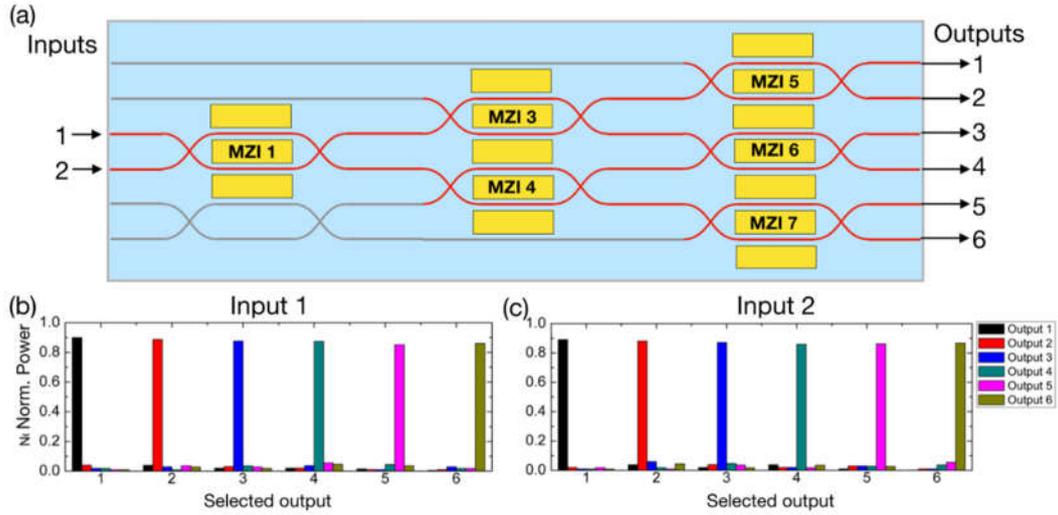

**Figure 4.** (a) Schematic of the 1 × 6 switch. Bar plots of the normalized output power of all 6 output ports when the chip is configured to switch light from input port 1 (b) and 2 (c) to the 6 output ports in an order from port 1 to port 6.

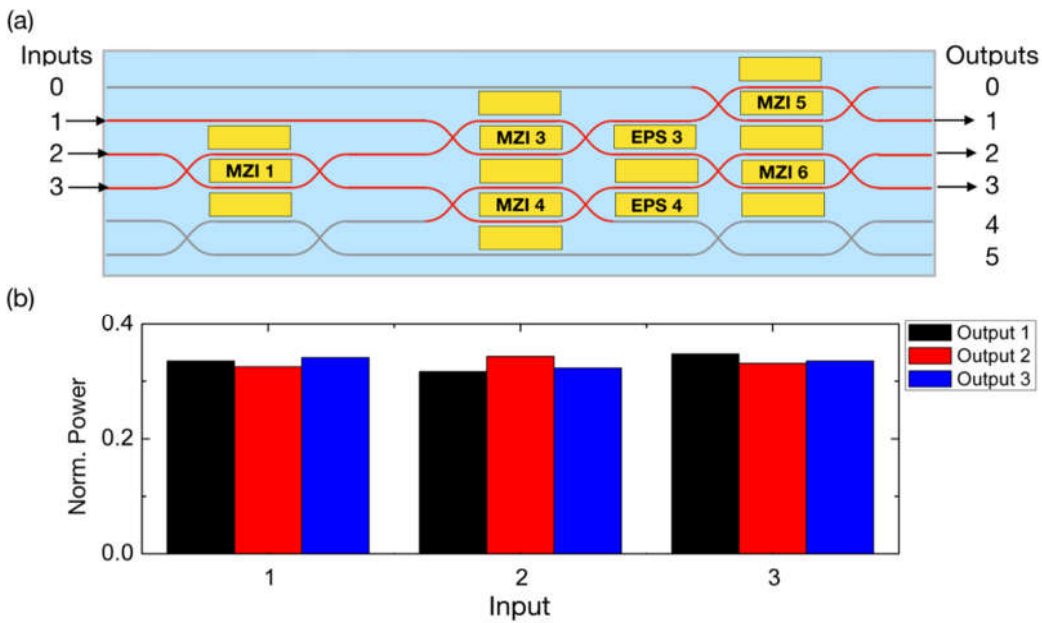

**Figure 5.** (a) Schematic of a balanced 3×3 interferometer. (b) Bar plots of the normalized output power of output ports 1 - 3 when the laser is sent into inputs 1 - 3.

*3.3 Balanced 3 × 3 interferometer*

Third, we demonstrate the balanced 3 × 3 interferometer using the waveguides highlighted in red in Figure 5a. Implementation of the same device has been demonstrated by Miller et al [16], in which the self-configuring approach used for realizing the balanced 3 × 3 interferometer has been described in great detail. We thus only present the measurement results in Figure 5b. One can see that by sending the laser beam into the balanced 3 × 3 interferometer from port 1, 2, and 3, the output powers from the three output ports are evenly splitted, namely, the output powers are close to one third of the total input power with deviations less than 5%.

## 4. Conclusion

To conclude, we have demonstrated realization and implementation of a reconfigurable multifunctional photonic integrated chip fabricated on LNOI substrate. Our device can be used as a high extinction ratio cascaded MZI, a 1 × 6 beam splitter, and a balanced 3 × 3 interferometer. The benchmark functionalities show that LNOI PIC devices can become a powerful platform for classic and quantum information processing owing to its unique advantages including the low propagation loss and high electro-optic coefficient.